\documentstyle[amssymb,multicol,epsfig,prl,aps]{revtex}

\begin{document}
\draft
 
\title{Jamming under tension in polymer crazes}

\author{J\"org Rottler and Mark~O.~Robbins}

\address{Department of Physics and Astronomy, The Johns Hopkins
University, 3400 N.~Charles Street, Baltimore, Maryland 21218}

\date{\today}

\maketitle

\begin{abstract} 
Molecular dynamics simulations are used to study a unique expanded
jammed state.  Tension transforms many glassy polymers from a dense
glass to a network of fibrils and voids called a craze.  Entanglements
between polymers and interchain friction jam the system after a fixed
increase in volume. As in dense jammed systems, the distribution of
forces is exponential, but they are tensile rather than compressive.
The broad distribution of forces has important implications for fibril
breakdown and the ultimate strength of crazes.
\end{abstract}


\begin{multicols}{2}
\narrowtext 

The common features of jammed systems ranging from molecular glasses
to granular media have sparked great
interest \cite{Liu2001,Ohern2001}. These systems jam as the available
volume becomes too small to allow relative motion of particles.
Unifying features of the jammed state are the presence of an
exponential force tail at large {\em compressive} forces and the
appearance of a network of forces on scales much larger than the size
of the constituent molecules or grains \cite{Liu1995,Coppersmith1996}.
The origins of these features are still debated.

\begin{figure}
\epsfig{file=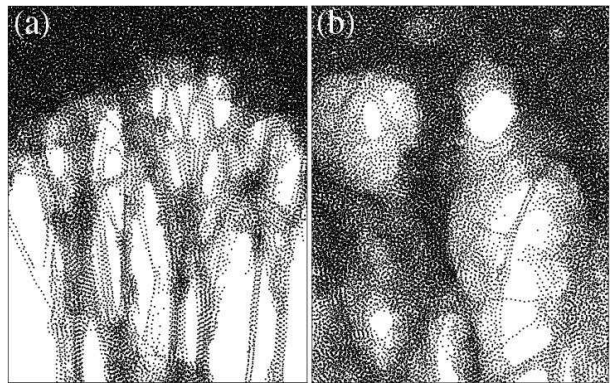,width=8cm}
\vspace{0.1cm}
\caption{Close-ups of interface between dense polymer and craze for
(a) $T=0.1 \epsilon/k_B$, $r_c = 1.5\sigma$ (fl) and (b) $T=0.3
\epsilon/k_B$, $r_c = 2.2\sigma$ (sfl). The lateral dimension of both
panels is $64\sigma$. The diameter and spacing of the fibrils increase
as $T$ and/or $r_c$ increase. }
\label{crazepic-fig}
\end{figure}

In this Letter, we consider a qualitatively different jammed state
that forms under {\em tension}.  Many amorphous polymers expand from a
typical dense glass to a ``craze'' consisting of a network of fibrils
and voids \cite{Kramer1990}. As shown in Fig. \ref{crazepic-fig}, the
dense and expanded jammed states coexist at a fixed tensile stress
$S$, much like liquid and gas phases coexist at a fixed pressure.
There are no covalent bonds that keep the craze from unravelling, but
experiments suggest that the topological constraints called
entanglements that limit dynamics in polymer melts behave like
crosslinking bonds \cite{Kramer1990}. These experiments cannot address
how deformation occurs, the distribution of forces within the system,
or the configurations of individual chains within the intricate craze
structure.  We describe extensive molecular dynamics simulations that
address all of these issues.  We find that the distribution of tensile
forces in the craze has an exponential tail analogous to that observed
for compressive forces in dense jammed systems.  The polymer undergoes
an approximately affine displacement as it deforms into the craze.
Expansion stops when segments that are only 1/3 of the distance
between entanglements are pulled taut.  This factor can be understood
from simple geometric arguments and the assumption that entanglements
act like chemical crosslinks between chains.

We study a standard coarse-grained model \cite{Puetz2000}, where each
linear polymer is modeled by $N$ beads of mass $m$. Van der Waals
interactions are modeled with a standard 6-12 Lennard-Jones (LJ)
potential with energy scale $\epsilon$ and length scale $\sigma$. A
simple analytic potential \cite{bondpot-comm,Sides2001} is used for
covalent bonds, with equilibrium length $l_0 = 0.96\sigma$ between
adjacent beads along the chain. In this fully flexible (fl) model, the
number of beads between entanglements (entanglement length) is
$N_e^{\rm fl} \approx 70$ \cite{Puetz2000}. In order to analyze
entanglement effects, we also consider semiflexible chains (sfl) where
an additional bond-bending potential stiffens the chain locally and
reduces the entanglement length to $N_e^{\rm sfl} \approx 30$
beads \cite{Sides2001,Faller2000,anglepot-comm}. We consider two
temperatures $T=0.1\epsilon/k_B$ and $T=0.3\epsilon/k_B$, where the
latter is close to the glass transition temperature. The strength of
adhesive interactions between beads is varied by truncating the LJ
force at either $r_c=1.5\sigma$ or $r_c=2.2\sigma$ \cite{Baljon2001}.
An amorphous glassy state is created in a cubic simulation cell of
side $L$ using standard techniques \cite{Puetz2000}. The period in the
$z$-direction is then increased at a small constant velocity
\cite{Baljon2001}. After an initial nucleation period, the system
separates into two phases (Fig.~\ref{crazepic-fig}).  The craze
network grows at constant tensile stress $S$ through plastic flow in a
narrow interfacial region called the ``active zone.''  The volume
occupied by the polymer increases by an ``extension ratio'' $\lambda$
that is remarkably insensitive to all parameters other than $N_e$.
For example, increasing the temperature and interaction range leads to
much larger fibrils in Fig.~\ref{crazepic-fig}(b) than (a), yet there
is no measurable change in $\lambda$. From the ratios of the initial
and final densities we obtain $\lambda_{\rm fl}=6.0 \pm 0.6$ and
$\lambda_{\rm sfl}=3.5 \pm 0.3$, independent of $N$, $T$, and $r_c$.
Experimental values of $\lambda$ range from about 2 to 7
\cite{Kramer1990}.

\begin{figure}[bt]
\epsfig{file=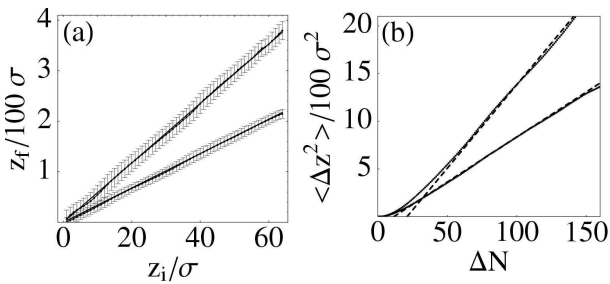,width=8cm}
\vspace{0.1cm}
\caption{(a) Final vs. initial height for flexible (large slope) and
semiflexible (small slope) chains ($T=0.1\,\epsilon/k_B$,
$r_c=1.5\sigma$). Averages were performed over z-intervals of width
$\sigma$. Straight lines have slope $\lambda_{\rm fl}=5.9$ and
$\lambda_{\rm sfl}=3.5$, respectively. Error bars show the standard
deviation from the averages at each height and are $17\pm
1\sigma$ (fl) and $9\pm 1\sigma$ (sfl). (b) Height change
$\Delta z$ as a function of distance $\Delta N$ from the chain
center. Straight lines have slope $\lambda^2l_pl_0/3$ with
$\lambda$ from (a). Systems at different $T,r_c$ and $N$ show
the same results.}
\label{affinedisp-fig}
\end{figure}

Although strain rate is strongly localized at the instantaneous position of
the active zone, the net effect is a nearly uniform or affine
expansion of each region of the polymer.  As shown in Figure
\ref{affinedisp-fig}(a), the final height $z_f$ of beads with an
initial height of $z_i$ is very close to $\lambda z_i$.  The rms
variation in the final height (half the errorbar length), has a
relatively small constant value. For reasons discussed below, it is
close to $N_e/3$ for both flexible and semiflexible chains at all $T$
and $r_c$ studied.  There are also much smaller rms displacements in
the $x-y$ plane.  These allow chains to lower their free energy by
gathering into fibrils whose local density is close to that of the
dense glass.  The magnitude of lateral displacements varies with the
size and spacing of fibrils, which depends on both $T$ and $r_c$.  For
example, the lateral displacement increases from $\sim 2.5\sigma$ in
Fig.~\ref{crazepic-fig}(a) to $\sim 5.6\sigma$ in
Fig.~\ref{crazepic-fig}(b).

\begin{figure}[bt]
\epsfig{file=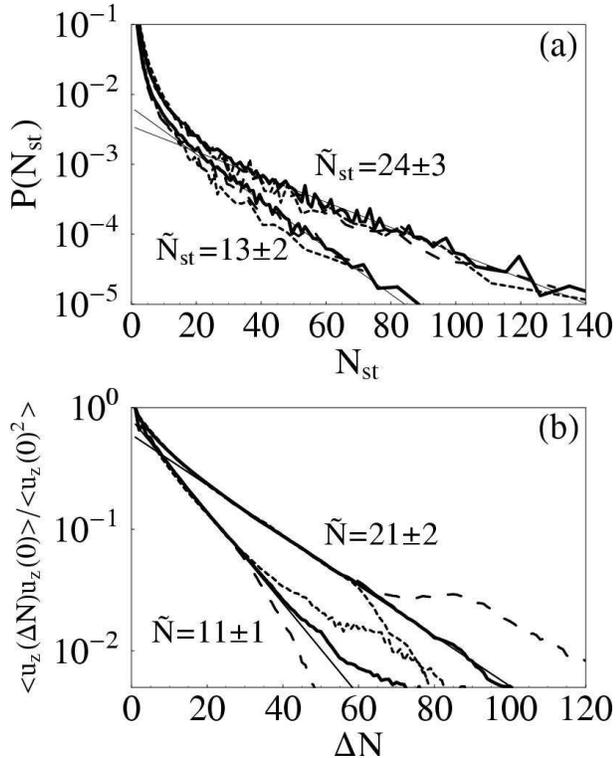,width=8cm}
\vspace{0.1cm}
\caption{(a) Probability of straight segments of length $N_{\rm st}$
for flexible and semiflexible chains.  Thick lines:
$T=0.1\,\epsilon/k_B$, $r_c=1.5\sigma, N=512$ and 1048576
beads. Dotted lines: $T=0.3\epsilon/k_B$, $r_c=2.2\sigma, N=512$ and
262144 beads. Long dashed lines: $T=0.1\epsilon/k_B$, $r_c=1.5\sigma ,
N=256$ and 262144 beads. (b) correlation function for the z-components
of the bonds for the same systems. Thin solid lines in both panels
show fits to an exponential decay with indicated decay lengths.}

\label{straightsegments-fig}
\end{figure}

The affine nature of the deformation along $z$ can also be inferred from the
change in the conformation of individual polymers.  In the initial
state, chains exhibit an ideal random walk (RW) structure inherited
from the melt.  The rms distance $\Delta r(\Delta N)$ between beads
that are $\Delta N$ neighbors apart scales as $\Delta r^2 =
l_pl_0\Delta N$, which defines the persistence length $l_p$. In the
dense glass, the mean-squared projection along each of the three axes
is equal.  An affine deformation by $\lambda$ along z will only change
the displacements along the $z$-axis, yielding $\langle\Delta
z^2\rangle=\lambda^2l_pl_0\Delta N/3$.
Fig.~\ref{affinedisp-fig}(b) shows that the final conformation of
chains follows this expression at large scales. The in-plane
components of the displacement are not changed during crazing.
The expansion along z pulls short segments of the chains taut, so that
$\Delta z^2$ rises quadratically at small $\Delta N$ in
Fig.~\ref{affinedisp-fig}(b).  The typical number of beads in a taut
segment, $\tilde{N}_{\rm st}$, can be obtained from the intersection
between the two asymptotic scaling forms: $(\tilde{N}_{\rm
st}l_0)^2=\lambda^2l_pl_0\tilde{N}_{\rm st}/3$, yielding
$\tilde{N}_{\rm st}=\lambda^2l_p/3l_0$.  Inserting the observed values
of $\lambda$ and $l_p$ ($l_p^{\rm fl}=1.65\sigma$ and $l_p^{\rm
sfl}=2.7\sigma$), we arrive at $\tilde{N}_{\rm st}^{\rm fl}\simeq
21\pm 4$ and $\tilde{N}_{\rm st} ^{\rm sfl}\simeq 12\pm 2$,
respectively.  The length of taut sections can also be determined by
direct analysis of the chain geometry.
Fig.~\ref{straightsegments-fig}(a) shows the distribution $P(N_{\rm
st})$ of the number $N_{\rm st}$ of successive beads whose
displacement continues upwards or downwards.  Here a bond is
considered up (down) if it is within 45$^\circ$ of the +z (-z) axis.
For both flexible and semiflexible chains, $P(N_{\rm st})$ has an
exponential tail with a characteristic decay length that, like
$\lambda$, is independent of $N$, $T$, and $r_c$.  The decay lengths,
$\tilde{N}_{\rm st}^{\rm fl}\sim 24\pm 3$ and $\tilde{N}_{\rm st}^{\rm
sfl}\sim 13\pm 2$, are in good agreement with the prediction from RW
statistics.  Essentially the same length scales appear in the decay of
the correlation function for the $z$-component of successive bonds,
Fig.~\ref{straightsegments-fig}(b), or the bond-angle correlation
function (not shown).

It is not surprising that all the above lengths are comparable to the
deviations from a purely affine deformation in
Fig.~\ref{affinedisp-fig}(a), since segments of length $\tilde{N}_{\rm
st}$ are stretched taut from their initial RW configuration.  However,
a very successful expression for $\lambda$ is usually derived by
assuming that the taut segments contain $N_e$ beads rather than $N_e/3$
\cite{Kramer1990}.
The maximum extension ratio, $\lambda_{\rm max}$, is defined
as the ratio between the fully stretched length $N_e l_0$
and the initial end-to-end distance of a RW of $N_e$ steps
\begin{equation}
\lambda_{\rm max} \equiv N_e l_0/ (l_0 l_p N_e)^{1/2} = \sqrt{l_0N_e/l_p}  .
\label{eq:lambda}
\end{equation}
Calculated values of $\lambda_{\rm max}$ are very close to values
of $\lambda$ measured in experiments and in our simulations where
$\lambda_{\rm max}=6.5\pm 0.5$ and $3.5\pm0.3$ for flexible and
semiflexible chains, respectively.
However, it was noted in early work that fully stretched chains
would actually yield an even larger extension ratio
because the initial end-to-end
vector of segments of length $N_e$ is randomly oriented \cite{Kramer1983}.
Since the
mean-squared projection along any direction is only 1/3 of the total,
$\lambda$ would be $\sqrt{3}\lambda_{\rm max}$ for fully stretched chains.
The observation that
$\lambda \approx \lambda_{\rm max}$ implies that
the average length of stretched segments is only $N_e/3$, as we have shown.
However those segments that are initially along the $z$-axis
are fully stretched when $\lambda=\lambda_{\rm max}$,
and it appears that these few segments are sufficient to prevent
further elongation.  Thus the factor of 1/3 results from the random
nature of the entangled network.  Note that a recent study of chains
with random crosslinks \cite{Barsky2000} is consistent with
Eq.~(\ref{eq:lambda}), while all segments between crosslinks were
fully stretched in a similar study of ordered networks
\cite{Stevens2001}.

A crosslinked system will jam when there is a network of fully
stretched covalent bonds, but it is less clear how entanglements can
lead to a jammed network. The entanglement length is typically
determined from the response of a polymer melt to a sudden strain
\cite{Puetz2000,Doi1986}. The resulting time-dependent stress shows a
broad plateau where the polymer has relaxed on short length scales,
but is unable to relax at larger length scales because
interpenetrating loops of polymer cannot pass through each other.  The
entanglement length corresponds to the typical length along the
backbone between these constraints.  The stress ultimately decays by
the slow diffusion of polymers along their length until their ends
pass through the loops and release all remaining constraints.
Expansion of the polymer into a craze is also limited by the inability
of polymer loops to pass through each other and interchain friction
prevents the chains from unraveling by the snake-like motion that
allows diffusion in the melt. The tension pulling two interpenetrating
loops in opposite directions creates large compressive forces
at the entanglements.  As in conventional jammed systems this increases
the barrier for sliding.
Long chains have many interpenetrating loops, and the tensions pulling in
each direction nearly cancel. Chains with $N<2N_e$ disentangle rather
than forming a craze \cite{Baljon2001}.

The distance between entanglements is determined by the chain
statistics \cite{Fetters1994} which are inherited from the melt.
They are not sensitive
to temperature because there is no evolution of large scale structure
in the glassy phase, and they are relatively insensitive to the
strength of adhesive interactions and density.  Thus $\lambda$ is
nearly constant, while the fibril structure at scales less than $N_e$
may vary dramatically (Fig. \ref{crazepic-fig}).  We have verified
that $\lambda$ only depends on chain statistics by artificially
varying the persistence length $l_p$ in the glassy state.
As expected, the value of $\lambda$ depends only on $l_p$ and not
on the flexibility of the chains, $r_c$, or other details of the
interaction potential.

\begin{figure}[hbt]
\epsfig{file=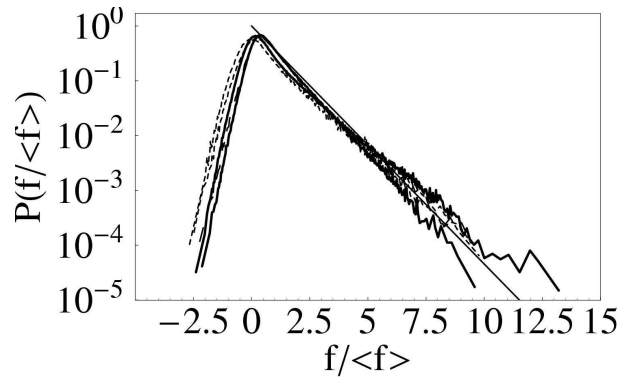,width=8cm}
\vspace*{0.1cm}
\caption{Distribution of tension $f$ in the craze for the systems of
Fig.~\ref{straightsegments-fig}. The straight line is
$\exp{(-f/\langle f \rangle})$. Only bonds under positive tension
were included in calculating $\langle f \rangle$.
$\langle f \rangle\approx 8.2
\epsilon/\sigma$ for $T=0.3\epsilon/k_B$, $r_c=2.2\sigma$, $\langle
f\rangle\approx 7.2\epsilon/\sigma$ in all other cases.}
\label{globaltension-fig}
\end{figure}

As in other jammed systems, the distribution of forces in the craze
follows a universal curve with an exponential tail
(Fig.~\ref{globaltension-fig}). The tensions $f$ in the covalent bonds
along the chains carry most of the stress. Only 10-20\% of covalent
bonds are under compression ($f<0$), and this part of the distribution
does not change much during crazing. The tensile ($f>0$) part of the
distribution is exponential, and results for all systems collapse
after normalizing by the average $\langle f \rangle$ over the tensile
region. Intuitively, one might expect the crazes with the larger
extension ratio (and $N_e$) to exhibit a larger average tension than
the less stretched crazes.  However, we find that $\langle f \rangle$
is a function of temperature $T$ and cutoff range $r_c$ only (see
caption) and does not depend on $N_e$. Expanding the craze elastically
beyond $\lambda$ increases $\langle f \rangle$, but the normalized
probability is unchanged until bonds begin to break.

In equilibrium, the energy of any region follows the exponential
Boltzmann distribution because this maximizes the number of
microstates available to the entire system at fixed total energy. In
jammed systems the total force is conserved, and one may imagine that
the exponential distribution of local forces also arises because it
maximizes the number of microstates. Two more detailed models for the
exponential distribution have been suggested for dense jammed systems.

The ``q-model'' \cite{Coppersmith1996} and its generalizations
\cite{Claudin1998} assume that stress propagates unevenly through the
system.  The fraction of stress passed to each neighbor, $q$, is
chosen at random from a distribution $R(q)$.  For a wide range of $R$
one obtains an exponential force distribution.  In addition, the
stress is concentrated in force chains like those seen in experiments
on granular systems.  The fibril network in a craze naturally provides
a branching path for transmission of stress.  Force is conserved along
each fibril and then redistributed at the nodes where fibrils merge or
split.  Study of the stress along chains shows that tension is
correlated along straight segments and then changes when the chain
changes direction at a node. A more detailed comparison to the q-model
is in progress.

Recently, O'Hern {\it et al.} \cite{Ohern2001} have proposed an
alternative explanation for the exponential force tail in jammed
materials. In equilibrium, the separation of beads at small distances
is weighted by a Boltzmann-factor $\exp[-V(r)/k_BT]$, where $V(r)$ is
the interaction potential.  The corresponding distribution of forces
is also nearly exponential if the force varies rapidly with $r$. The
repulsive part of the LJ potential satisfies this criterion, but our
covalent bond potential does not. We have calculated the distribution
of bond energies in the craze and find that it is not exponential.

The presence of an exponential stress distribution has consequences
for the ultimate fracture of craze fibrils. In a minimal model often
employed in the literature \cite{Jones1999}, the maximum breaking
force $F_{\rm max}$ of a fibril composed of $n$ strands is estimated
by assuming that each strand carries an equal share of the load and
all covalent bonds break at a critical force $f_c$. This implies
$F_{\rm max}/f_c=n$, and all strands break at the same time.  If the
distribution of forces among strands is exponential, the most stressed
strand will break at a lower $F_{\rm max}$.  The force needed to
initiate failure can be estimated using a simple scaling argument.
The first strand will break when
$nP(f>f_c)=n\int_{f_c}^{\infty}1/\langle f\rangle \exp{(-f/\langle f
\rangle)df} =n\exp(-nf_c/F_{\rm max})=1$, where we have used $\langle
f\rangle=F_{\rm max}/n$. This implies a breaking force of $F_{\rm
max}/f_c=n/\ln (n)$ instead of $n$. The redistribution of load after a
strand breaks will cause the remaining strands to break sequentially.
Note that no chain scission occurs in the simulations presented here,
but it does occur upon further straining of the craze.  A detailed
analysis of fibril breakdown will be presented elsewhere.

In summary, we have shown that crazing transforms a conventional dense
jammed state into a unique expanded jammed state. Expansion is limited
by the same interlocking of polymer loops that leads to entanglements
in polymer melts. However, the average length of straight segments is
only $N_e/3$ due to the random nature of the network. Interchain
friction prevents disentanglement of the loops. As in dense jammed
systems, the distribution of forces is exponential. However, the
forces are tensile rather than compressive.  We hope that contrasting
this new jammed state with conventional ones will help unravel the
microscopic origin of their common features.

This work was supported by the Semiconductor Research Corporation
(SRC) and by the NSF grant No. DMR0083286. We thank
E.~J.~Kramer,H.~R.~Brown, and C.~Denniston for useful discussions.

\end{multicols}

\begin{references}
\bibitem{Liu2001} A.~J.~Liu and S.~R.~Nagel (Eds.), {\it
Jamming and Rheology}, (Taylor \& Francis, London, 2001)
\bibitem{Ohern2001} C.~S.~O'Hern, S.~A.~Langer, A.~J.~Liu,
and S.~R.~Nagel, Phys.~Rev.~Lett. {\bf 86}, 111 (2001)
\bibitem{Liu1995} C.-h.~Liu, S.~R.~Nagel,
D.~A.~Schecter, S.~N.~Coppersmith, S.~Majumdar,
O.~Narayan, T.~A.~Witten, Science {\bf 269}, 513 (1995)
\bibitem{Coppersmith1996} S.~N.~Coppersmith, C.-h. Liu, S.~Majumdar,
O.~Narayan, and T.~A.~Witten, Phys.~Rev.~E {\bf 53}, 4673 (1996)
\bibitem{Kramer1990} E.~J.~Kramer, L.~L.~Berger, Adv. Polymer
Science {\bf 91/92},1 (1990).
\bibitem{Puetz2000} M.~P\"utz, K.~Kremer, G.~S.~Grest,
Europhys.~Lett. {\bf 49}, 735 (2000).  The quoted $N_e$ is from the
plateau modulus.
\bibitem{bondpot-comm} The potential is $V_{\rm
br}(r)=-k_1(r-r_c)^3(r-R_1)$, and the constants $k_1$ and $R_1$ are
adjusted to fit the equilibrium bond length and to allow for bond
breaking when the tension exceeds 100 times the breaking force $f_{\rm
LJ}$ of the van der Waals bonds.
\bibitem{Sides2001} S.~W.~Sides, G.~S.~Grest, M.~J.~Stevens,
Phys.~Rev. E {\bf 64}, 050802 (2001), Macromolecules {\bf 35}, 566
(2002).
\bibitem{Faller2000} R.~Faller, F.~M\"uller-Plathe, and A.~Heuer,
Macromolecules {\bf 33}, 6602 (2000) and R.~Faller and
F.~M\"uller-Plathe, ChemPhysChem. {\bf 2 }, 180 (2001).
\bibitem{anglepot-comm} The bond-bending forces are modeled with
$V_{B}=\sum_{i=2}^{N-1}b\left(1-\frac{(\vec{r}_{i-1}-\vec{r}_{i})\cdot
(\vec{r}_{i}-\vec{r}_{i+1})}{|(\vec{r}_{i-1}-\vec{r}_{i})||(\vec{r}_{i}-
\vec{r}_{i+1})|}\right)$, where $\vec{r}_{i}$ denotes the position of
the $i$th bead along the chain, and $b$ characterizes the stiffness.
\bibitem{Baljon2001} A.~R.~C.~Baljon, M.~O.~Robbins, Macromolecules
{\bf 34}, 4200 (2001).
\bibitem{Kramer1983} E. J. Kramer, Adv. Polymer Sci. {\bf 52/53} 1 (1983).
\bibitem{Barsky2000} S.~Barsky and M.~O.~Robbins, unpublished
\bibitem{Stevens2001} M.~J.~Stevens, Macromolecules {\bf 34}, 1411
(2001)
\bibitem{Doi1986} M.~Doi and S.~F.~Edwards, {\it The Theory of Polymer
Dynamics}, (Oxford University Press, Oxford, 1986)
\bibitem{Fetters1994} L.~J.~Fetters, D.~J.~Lohse, D.~Richter,
T.~A.~Witten, and A.~Zirkel, Macromolecules {\bf 27}, 4639 (1994)
\bibitem{Claudin1998} P.~Claudin, J.-P.~Bouchaud, M.~E.~Cates, and
J.~P.~Wittmer, Phys.~Rev.~E {\bf 57}, 4441 (1998)
\bibitem{Jones1999} R.~A.~L.~Jones, R.~W.~Richards, {\it Polymers at
Surfaces and Interfaces}, (Cambridge University Press, Cambridge,
1999).

\end{references}
\end{document}